\documentstyle[aps]{revtex}
\begin{document}

%TCIDATA{TCIstyle=Article/art2.lat,aps,revtex}

\title{Vertex Corrections in nearly Ferroelectric
Superconductors}
\author{D. Fay$^1$ and M. Weger$^2$}
\address{$^1$I. Institut f\"ur Theoretische Physik,
Universit\"at Hamburg, Jungiusstr. 9, 20355 Hamburg, Germany}
\address{$^2$Racah Institute of Physics, Hebrew University,
Jerusalem, Israel}
\date{\today}
\maketitle
\begin{abstract}
We investigate the effect of an incipient ferrolectric transition
on vertex corrections to the superconducting pairing interaction.
The vertex corrections for small momentum transfers are large
independent of the type of Boson responsible for the 
superconducting transition. The electron-phonon interaction is
found to be enhanced by a nearly ferroelectric medium. We 
discuss application to the cuprate superconductors.
\end{abstract}
\vspace{0.5in}
1) INTRODUCTION.

High-temperature superconductivity is commonly believed to be
associated with the exchange of antiferromagnetic spin fluctuations.
A unified description of antiferromagnetism and
superconductivity is given by a SO(5) group scenario \cite{Zhang}. 
However, it had been previously shown by 
Birman and Solomon \cite{Birman} that a similar group-theoretical 
formulation is possible
with superconductivity associated with charge-density waves,
ferromagnetism, and ferroelectricity, as well as the "common" 
antiferromagnetic scenario. We consider here an effect associated 
with the nearness of a ferroelectric transition. A phenomenological
pairing interaction has been proposed \cite{Weger_etal,Peter} for 
such a nearly ferroelectric system and one purpose of this paper 
is to make a first step towards giving this pairing interaction a 
firmer microscopic basis.

Quite generally,  the pair interaction can be viewed as the
exchange of a Boson between the members of a Cooper pair.  Of
primary interest here is 
the electron-Boson vertex.  We re-evaluate the Bardeen calculation
\cite{Bardeen} of the correction to the electron-phonon vertex due
to electron-electron interactions, for the case of an electron-gas
imbedded in a dielectric medium with  a large dynamic dielectric 
constant $\varepsilon$ reflecting the nearness to a ferroelectric
transition. We calculate the lowest order vertex correction
that includes phenomenologically the important ionic dielectric
effects. This turns out to be a large correction. Although we
assume in the following that the Boson is a phonon,  it is
important to note that the present result is not limited to this
case and indeed would also yield a large vertex correction in the
popular scenario in which the Boson is an antiferromagnetic
spin fluctuation. 

Many perovskites are ferroelectrics. Perovskites like SrTiO$_3$ 
and the high-T$_c$ cuprates are nearly ferroelectric; their ionic
dielectric constant is abnormally high, although they do not
undergo a ferroelectric transition. The dielectric constant of
La$_{2-x}$Sr$_x$CuO$_4$ and YBa$_2$Cu$_3$O$_{7-\delta}$  
was measured recently quite accurately as function of frequency 
and is of the order of 100 at frequencies of order 10 meV 
\cite{Kircher,Henn}. In ''conventional'' calculations of the 
electron-phonon interaction, this dielectric constant is not taken 
into account. The reason is that, although  the electron-phonon 
interaction itself is a nonadiabatic contribution, it is 
{\it calculated} within the framework of the Born-Oppenheimer 
approximation (BOA); i.e., it is assumed that during the 
scattering of the electron, while it imparts momentum to the 
lattice, the lattice does not move. In contrast, the ionic 
dielectric constant describes motion of the lattice. Therefore, 
in order to take it into account, we must consider effects 
outside the BOA. This presents a {\it fundamental } difficulty, 
and not just ''technical'' considerations.

The phonon-mediated interaction considered by Bardeen and Pines
in the 1950's involves an effect outside the BOA. The scattering
of electron 1 causes an atomic motion, and this motion interacts 
with electron 2. The Bardeen-Pines interaction is given by :
V = g$^2\frac{2\Omega }{\omega^2-\Omega ^2}$, where $\Omega $ 
is the phonon frequency, g is the Frohlich constant given by 
g = I/$\sqrt{M\Omega },$  M is the ionic mass, I is the
matrix element given by I(q) = $\langle k\mid \nabla V\mid k+q
\rangle $ , and V is the electron-ion potential. I(q) is thus
calculated {\it within} the BOA. This is an apparent paradox; 
this procedure is justified by Migdal's theorem which states 
that the corrections to I(q), arising from the fact that we consider 
an effect outside the BOA, are of order
$\Omega /E_F$ $\simeq $ $\sqrt{m/M}\simeq $ 10$^{-2}$ , 
where m is the electronic band mass.

Pietronero $et$ $al$ \cite{Pietr} suggest that in the high-T$_c$
cuprates $\Omega /E_F$ is not so small, and as a result the
deviations from Migdal's theorem are significant. Here, we also
question the validity of the BOA, but for a different reason - 
namely, we suggest that the very large ionic dielectric constant
has a very large effect on the electron-phonon matrix element 
I(q). It renormalizes it, and thus makes it frequency dependent; 
at very low frequencies I(q) is increased by a significant 
amount above the BOA value.%
\vspace{0.2in}

2) CONVENTIONAL\ APPROACH.

The ''conventional'' way to calculate the electron-phonon matrix 
element I(q) is based, first, on Bloch's calculation of 1928
\cite{Bloch}, in which he assumed:
I(q) = $\langle k\mid \nabla V\mid k+q\rangle $ , where V is the
electron-ion potential Ze$^2/r$ . Thus I(q) = 4$\pi Ze^2/q$ . 
I(q) is seen to diverge at small q values. We denote this as
the ''Bloch Vertex'' $\Gamma _{Bloch}$ (Fig.1a).  This estimate
neglects the electron-electron interaction which screens the
electron-ion potential. This screening was first considered by
Bardeen in 1937 \cite{Bardeen}. The electron-electron 
interaction propagator is given by:  
D$_{ee}$ (q) = 4$\pi e^2/q^2$ .  Bardeen took it into account
by considering the potential:  
\begin{equation}
V_{Bardeen}(q) = \frac{4\pi Ze^2/q^2}{%
1+D_{ee}(q)N(E_F)} = \frac{4\pi Ze^2}{q^2+q_D^2}, 
\end{equation}
where $q_D^2 = 4\pi e^2N(E_F)$.
This potential no longer diverges as q$\rightarrow$ 0. The
electron-phonon matrix element is given by:
I(q) = $\frac{4\pi Ze^2\mid q\mid }{q^2+q_D^2}$ .
Since q$_D$
%TCIMACRO{\TEXTsymbol{>} }
%BeginExpansion
\mbox{$>$}
%EndExpansion
k$_F$ , the dependence of I(q) on q is weak and the 
electron-phonon scattering is nearly isotropic \cite{Allen}. 
The Cooper potential reflects this isotropic nature and 
leads to superconducting pairing with s-wave symmetry. 
What Bardeen did to deal with this (without saying it 
explicitely) was to sum-up RPA bubble diagrams of the
form of Fig 1b to get the correct screened vertex. We 
call this the ''Bardeen vertex'' $\Gamma _{Bardeen}$ . 

The simplest vertex correction arising from the electron-electron
interaction, $\Delta\Gamma _{Coulomb}$, is shown in Fig. 1c. 
We calculate this contribution in the next section for comparison 
with the vertex correction including ionic effects. In Fig. 1c we 
could of course replace D$_{ee}$ with a screened interaction 
of the form of Fig. 1b. Note that Migdal's theorem does {\it not} 
apply here since D$_{ee}$ does not possess a low-energy 
cutoff.%
\vspace{0.2in}

3) VERTEX\ CORRECTION\ INCLUDING\ THE\ IONIC\ DIELECTRIC\ FUNCTION

In a medium with a dielectric constant $\varepsilon ,$ the
electron-electron interaction is given by:
$4\pi e^2/q^2\varepsilon $ . Here $\varepsilon $ is the dielectric
constant of the medium {\it external }to the electron gas
\cite{Gersten}. Thus, we use a dressed electron-electron 
propagator:
\begin{equation}
\tilde{D}_{ee} (q,\omega ) = \frac{4\pi e^2}{(q^2+q_{D}^2)
\varepsilon (q,\omega )} .
\end{equation}
Diagramatically, we denote $\tilde{D}_{ee}$ by a heavy broken
line (Fig.1d) that denotes the inclusion of the effect of the 
highly polarizable ions on the electron-electron interaction.
The Thomas Fermi screening vector $q_{D}$ is included to
facilitate the numerical computations and to account for
electronic screening.  Thus we consider the vertex correction
$\Delta \Gamma $ shown in Fig.1e. For simplicity in the
present calculation we take $\Gamma _{Bardeen} $ to be a
constant. We can thus write 
$\Gamma $ = $\Gamma _{Bardeen} \lbrack 1 + (\Delta \Gamma /%
\Gamma _{Bardeen})\rbrack $ and we calculate $\Delta \Gamma /%
\Gamma _{Bardeen}$ and $\Delta \Gamma_{Coulomb}/%
\Gamma _{Bardeen}$ which are to be compared with the 
number 1.

We have to calculate the vertex correction as a function of q,
paying particular attention to {\it small} q values. This is
plausible for several reasons. First, the ionic dielectric
corrections could also be included in the Bardeen screening
by replacing D$_{ee}$ with $\tilde{D}_{ee}$ and setting
 $q_{D}=0$ to avoid double-counting of screening diagrams.
Then if $\tilde{D}_{ee}$ were zero (as is the case for an
infinite $\varepsilon),$ $\Gamma $ would be the Bloch vertex
which diverges as q$\rightarrow 0.$ Thus
$\Delta \Gamma /\Gamma _{Bardeen}$  would diverge
as q$\rightarrow 0.$
 
The necessity to consider small q values was also pointed out
by Pietronero $ et$ $al$ \cite{Pietr}, for a different reason.
They show that for their (phonon) vertex correction, the
behavior when q$\rightarrow 0$ first, and $\omega
\rightarrow 0$ afterwards, is entirely different from the
behavior when $\omega \rightarrow 0$ first, and
q$\rightarrow 0$ afterwards, which was previously calculated
by Grabowski and Sham \cite{Sham}.

The relevance of small q values for the cuprates has also 
been pointed out previously by Santi $et$ $al$ \cite{Santi}, 
Perali and Varelogiannis \cite{Varelog},  Abrikosov 
\cite{Abrik}, Zeyher and Kulic \cite{Zeyher}, Bulut and 
Scalapino \cite{Bulut}, Tsuei and Kirtley \cite{Tsuei}, Leggett 
\cite{Leggett}, and others. The observation of stripes 
\cite{Bianconi} suggests the relevance of q values close 
to q $_{stripe}$ $\simeq 0.24$ A$^{-1}$ \cite{Rome} and 
we calculate here the vertex correction for q values in 
this range. %
\vspace{0.2in}

The dielectric constant has a sharp dispersion at a frequency
of about 19 meV in YBCO \cite{Kircher}. This frequency is the
frequency of phonons involving the displacements of Barium
atoms in the c-direction. In LSCO , the dispersion frequency
is about 27 meV, the frequency of phonons involving motion
of strontium in the c-direction \cite{Henn}. A theoretical
expression for the ionic dielectric constant is given by the
Lyddane-Sachs-Teller theory (LST) \cite{LST}, namely:
\begin{equation}
\varepsilon (\omega ) = \frac{\omega ^2-\omega _L^2}{\omega ^2-\omega
_T^2}\, \varepsilon _\infty ,
\end{equation}
where $\omega_L$ and $\omega_T$ are the frequencies of the
longitudinal and transverse phonon modes. Although this expression
is derived for q=0, we assume that it is valid for the small values
of q of interest here. 

The use of such a simplistic expression, however,  
raises several further questions:

(a) There are several phonon modes, and there is no $a-$ $priori$
reason why $\varepsilon (\omega )$ should be dominated by one
transverse, and one specific longitudinal mode.

(b) We ignore the dispersion of the phonon modes for the 
relatively small values of q that we consider here (about 
1/4-1/3 of the way to the Brillouin zone). This is justified 
here since there is no softening of the modes, as 
manifested by absence of a strong temperature 
dependence of their frequency. \cite{Appel}

(c) This expression applies to insulators. The cuprates possess 
a metallic Fermi surface.

(d) The expression is derived for an isotropic 3D system.

We feel reasonably confident using Eq.(3) simply because 
experiment shows that it works quite well. \cite{Kircher,Henn}.
For example, from optical infrared measurements, it is found that 
the simple LST expression applies extremely well to the c-axis 
component of $\varepsilon $ for LaSrCuO 
(with $\omega _L$ =63 meV $) \cite{Henn},$ and
reasonably well for YBCO \cite{Kircher}; (there are two modes 
that serve as $\omega _L$ , at 40 meV and 70 meV; we can 
choose some average in-between these two frequencies; 
about 50 meV gives good agreement, with 
$\varepsilon _\infty $ $\simeq 4).$

Since the ''metallic'' layer is thin, the width being about 2a$_c$
where a$_c $ is the Bohr radius of the oxygen 2p$\sigma $
orbitals in the c-direction, a$_c$ $\simeq $ 0.4 $\AA$, and the 
average distance between electrons is approximately the 
lattice constant  ($\simeq 4$ $\AA$), the c-axis component of 
$ \varepsilon $ screens out the electron-electron interaction 
effectively, and we can use it (as determined experimentally) 
in the expression for $\tilde{D}_{ee}$.

The contribution $\Delta\Gamma$,  Fig. 1e,  is given at zero
temperature by
\begin{equation}
\Delta\Gamma(\underline{k},\underline{q}) = 
\Gamma_{Bardeen}\sum_{\bf p}\int\frac{dp_0}{2\pi}
\tilde{D}_{ee}(\underline{p}-\underline{k})G_0(\underline{p})
G_0(\underline{p}+\underline{q}),
\end{equation}
where $\underline{k}=(\bf k$,$k_0)$ and we assume a two 
dimensional system. Substituting  Eq. (3) for 
$\varepsilon (\omega )$ with $\varepsilon_\infty = 1$  in 
Eq.(2) we have
\begin{equation}
\tilde{D}_{ee}(\underline{k}) = \frac{4\pi e^2}{({\bf k}^2+q_{D}^2)}
\frac{k_0^2 - \omega_T^2}
{k_0^2 - \omega_L^2 + \imath\delta}.
\end{equation}
This form is quite plausible but, as mentioned above, has not 
yet been derived from a truly microscopic theory. $G_0$ is 
the usual zero order electron propagator and we assume for 
now a free particle spectrum.  The $p_0$ integration can be 
caried out as a contour integration picking up contributions 
from the poles of the $G_0$'s and $\tilde{D}_{ee}$.

Since $\Gamma$ is normally part of a larger diagram, the pair
interaction, for example, its external momenta and energies are
integrated over.  Although for such applications we require the
full complex $\Delta\Gamma$ as a function of all of its
arguments, for present purposes it should suffice for a rough
estimate of $\Delta\Gamma$ as a function of 
$q\equiv \vert\bf q\vert$ to take typical values for the other 
variables.  We also neglect for now the imaginary part of 
$\Delta\Gamma$ .  We assume $0<q_0<\omega_L$ and, 
because the external electron lines usually represent 
members of a Cooper pair, we take
$\vert\bf k \vert$=$k_F$ and $k_0=0$, measuring energies 
relative to $E_F$. Thus we have
\begin{equation}
\Delta\Gamma/ \Gamma_{Bardeen}=\frac{1}{2\omega_L}\int\int
\frac{dp_xdp_y}{(2\pi)^2}
\frac{4\pi e^2} { ( {\bf p} - {\bf k} )^2 + q_{D} ^2 } 
iI({\bf p} ,q, {\bf p}\cdot{\bf q} ;\omega_T,\omega_L),
\end{equation}
where $\bf k$ = $k_{F}\bf\hat{k}$ and we have taken unit volume and 
set $\hbar = 1.$ I consists of six terms resulting from the $p_0$
integrations which are displayed in the Appendix. For the 
p integrations we can set the $p_x$ axis 
along the direction of $\bf q$ and then the direction of $\bf k$ 
with respect to  $\bf q$ is given by $\hat{k}_x = \cos(\theta_k)$. 
For the present computation we have taken as typical values
$q_D = 0.1k_F$, $\hat{k}_x = 0.8$, $\omega_L=0.16E_F$, and
$\omega_T = 0.04E_F$.  

For diagram 1c the complications due to the ionic corrections are
absent and only two terms result from the $p_0$ integrations.
Then $\Delta\Gamma_{Coulomb}/\Gamma_{Bardeen}$ is given by
Eq.(6) without the factor $1/2\omega_L$ and with I replaced by 
$I_{Coulomb}$ where
\begin{equation}
iI_{Coulomb} = \frac{f(\varepsilon_p)[1-f(\varepsilon_{p+q})]}
{\varepsilon_{p+q}-\varepsilon_p - q_0 - i\delta} - 
\frac{f(\varepsilon_{p+q})[1 - f(\varepsilon_p)]}
{\varepsilon_{p+q}-\varepsilon_p - q_0 + i\delta}
\end{equation}
and $f(\varepsilon_p)$ is the Fermi function at $T=0.$ 
The principal value integrations over $p_x$ and $p_y$ must be
done numerically.

In Fig. 2 we show
$\Delta\Gamma/\Gamma_{Bardeen}$ and
$\Delta\Gamma_{Coulomb}/\Gamma_{Bardeen}$ as functions of
$q/k_F$ for $q_0 = 0.08E_F$ (solid curves) and 
$q_0 = 0.02E_F$ (dashed curve).  There are two surprising 
results here. First, in 2D the vertex correction 
$\Delta\Gamma_{Coulomb}/\Gamma_{Bardeen}$ is already 
very large and, second, the effect of the ionic screening is 
drastic: One would naively expect the inclusion of the large 
ionic dielectric function to greatly reduce the correction. 
Except very near the sign change just below $q=0.3k_F$, 
this is {\it not} the case. A similar sign change was also found 
by Pietronero $ et$ $al$ \cite{Pietr}. 

The structure in $\Delta\Gamma_{Coulomb}/\Gamma_{Bardeen}$
deserves a brief comment. It arises from the complicated structure
of the integrands and does not seem to have any obvious physical
relevance. It is due to the first term in Eq. (7) which is singular at 
${\bf p}={\bf p_s}(q,q_0)$ when 
$q_0=\varepsilon_{p_s+q}-\varepsilon_{p_s}$,  i.e., when the
external frequency can excite a particle-hole pair. The integrand 
of the principal value integration in Eq. (6) then contains the 
singularity at $\bf p=\bf p_s$ and a fixed (for $\bf k$ fixed) peak 
arising from the Coulomb factor. As q changes (for fixed $\bf k$ 
and $q_0$), the singularity moves through the Coulomb peak 
causing alternately positive and negative contributions,  
producing the structure. This structure does not occur in 
$\Delta \Gamma / \Gamma _{Bardeen}$ because the 
integrand on the negative side of the singularity becomes 
positive a short distance from the singulatity and practically 
no negative contribution occurs.%
\vspace{0.5in} 

4) THE\ LARGE\ VALUE\ OF\ $\Delta \Gamma /\Gamma _{Bardeen}.$

The extremely large value of $\Delta \Gamma /\Gamma _{Bardeen}$
that we find here seems at first sight to be unphysical, since 
(to the  best of our knowledge) it was not considered before. We 
suggest that it {\em is} physical. We are assuming here that our
lowest order correction is not too strongly reduced by higher 
order vertex corrections, see discussion below. We believe that the 
reason why such an effect was not considered previously, is that 
in a homogeneous system, the dielectric constant $\varepsilon $ 
also enters into the expression for the bare vertex, which 
becomes: \\
Ze$^2$ $\mid q\mid /q^2$ $\varepsilon $ , and not 
only into the expression for $\tilde{D}_{ee}$ . Our ansatz 
considers an extremely inhomogeneous system 
\cite{Rome,Weger_etal,Peter}.

While we consider effects due to phonons, we consider 3 entirely
distinct phonon modes. Namely, the mode $\Omega=q_0$ which
represents phonons associated with the momentum q of the bare 
vertex $\Gamma _{Bloch}$, Fig. (1a); and
the modes $\omega _T$ and $\omega _L$ , which dominate the
ionic dielectric constant. Since the modes are distinct, they
involve motion of different atoms, in different regions of the
unit cell. As mentioned previously, the mode $\Omega $ 
could also be a spin fluctuation (or another type of Boson). For
now we assume it is a phonon involving motion of the planar 
oxygen; such motion could be longitudinal (along the Cu-O bond) 
or transverse, in the a-b plane or in the c-direction. The frequency 
of the transverse motion is around 40 meV. This mode is seen as 
perhaps a McMillan-Rowell structure in tunneling experiments in
Bi$_2$Sr$_2$CaCu$_2$O$_{8+\delta}$ \cite{Vedeneev}.
The mode $\omega _T$ involves the motion of the Ba (or Sr) atoms
in the c-direction. The mode $\omega _L$ involves motion of the
apex oxygen, as well as motion of the planar oxygens. These modes
are in the range 15-80 meV; $\omega _T$ is in the range 15-30 meV, 
and  $\omega _L$ is in the range 40-80 meV. This distinction between
the modes is important not only to avoid double counting, but also 
because it causes the shielding of the electron-electron interaction 
D$_{ee}$ by the ionic dielectric constant to be {\it entirely different}
from the shielding of the electron-$\Omega $ phonon interaction
(by the ionic dielectric constant, associated with the modes
$\omega _T$ , $\omega _L).$ We estimated this last shielding and
found that it is small \cite{Weger_etal}. Therefore we do not
introduce it into the present calculation.

While the experimental determination of $\varepsilon (\omega )$ by 
the IR measurements \cite{Kircher,Henn} is definitive, it is 
instructive to consider possible microscopic causes for the 
anomalously large value of \\ $\varepsilon $ . We believe that the 
large $\varepsilon $ may be related to a new degeneracy between 
the Zhang-Rice singlet and the anti Jahn-Teller triplet of the
CuO$ _5 $ complex. Kamimura $et$ $al$ \cite{Kamimura} 
calculated the splitting between the singlet and triplet as function 
of the occupation of the 3d shell of the {\em chain }copper, and 
found that for 0.55 holes there, these states are degenerate. This 
calculation is carried out using a quantum-chemistry algorithm. 
We can characterize this splitting by an effective Hubbard U, and 
write: U$_{eff}$ = U$_{bare}/\varepsilon $ . U$_{bare}$ is the 
Hubbard U in a ''normal'' complex, $i.$ $e.$ several electron-volts. 
Thus the splitting of about one tenth of an eV calculated by 
Kamimura, indicates a value of $\varepsilon $ of about 50. Also, 
this calculation shows clearly the {\em local} nature of this large
dielectric constant.

Anisimov $et$ $al$ \cite{Anisimov} found that the singlet and 
triplet are close, by a rigorous LDA calculation. They also
considered relaxation of the complex, $i. $ $e.$ different
Cu-O distances in the singlet and triplet states. They found 
a rather large relaxation, substantiating the ionic nature of
the large dielectric constant. Stern $et$ $al$ \cite{Polinger} 
found a near-degeneracy of the singlet and triplet states, 
near the Sr in LSCO by EXAFS measurements. 
Itai \& Gatt \cite{Gatt} carried out a quantum chemistry
calculation, and showed the important role of the c-axis 
motion of the alkaline-earth atom. All these calculations show 
that the $\varepsilon ^{\prime }s$ that describe the shielding 
of the electron-electron interaction D$_{ee}$ and the 
electron-planar oxygen interaction $\Gamma _{Bloch}$ are 
of an entirely different nature, and only the former one is 
greatly enhanced. Therefore the very large difference in 
their values is not ''unphysical''.

Since the diagram, Fig. 1e yields such a large result one can not
rule out that further vertex corrections may also be important.
In fact, the natural class of diagrams to include is the sum of
"ladder" diagrams. Such diagrams represent nothing more than
the spin susceptibility, or spin fluctuation propagator, and
describe the (antiferromagnetic) spin correlations thought to be
particularly important in the cuprate superconductors. Previous
calculations \cite{Manske,Kim}, which considered the effect of
these contributions on the phonon frequencies and on the
screening of the electron-phonon vertex in nearly
antiferromagnetic systems, could be modified to include the
ionic screening considered here. Thus the nearness of both a
ferroelectric as well as an antiferromagnetic transition could
perhaps be accounted for.

   Kim \cite{Kim} first considered effects of the proximity to
a ferromagnetic or an antiferromagnetic transition, and showed
that this can enhance the electron-phonon coupling constant.
The enhancement arises both from softening of the phonon,
and from an increase of the matrix element
$\langle k\mid \nabla V\mid k+q\rangle $ . The first effect
does not concern us here, (it doesn't lead to an increase in
$T_c$ in the strong-coupling case). The second effect bears
some similarity to the effect that we consider here, except
that we consider primarily the nearness to a ferroelectric
transition. 

   Kim considers the role of a sum-rule that restricts the
magnitude of his effect. This sum-rule was also considered by
McMillan \cite{McMill} and originally proved by Heine, Nozieres, 
and Wilkins \cite{Heine}. It states that $V(0)$ = Z/$N(E_F)$ .
This rule applies to a {\it homogeneous} system. In an
inhomogeneous system, it breaks down. Its origin is that when
$\tilde D_{ee}$ is reduced from $D_{ee}$ by $\varepsilon$,
the bare vertex $\Gamma$ is also reduced by $\varepsilon$
(Fig. 1b) and these two effects cancel each other when q=0.
(When $q>0$, a large $\varepsilon$ actually diminishes V(q)). 

   In an inhomogeneous system, the reduction of $D_{ee}$ 
and the reduction of $\Gamma$ are given by two different 
dielectric constants. We showed \cite{Rome,Weger_etal,Peter}
that the $\varepsilon$ for $\Gamma$ is close to one, while the
$\varepsilon$ for $D_{ee}$ is large. This follows from the 
$\it local$ nature of $\varepsilon$  \cite{Roehler,Kamimura,Peter}.

   We note in passing that the relevant values of q are on the 
order of 0.25$\AA^{-1}$. \cite{Rome} This is small enough that
the use of the $\varepsilon$ derived for q=0  is justified. On the 
other hand, the screening length $q_{D}^{-1}$ is about 4$\AA$ 
which is small enough so the dielectric constant is local.
\vspace{0.5in} 

5)  BARDEEN'S\ VERTEX\  FOR\ AN\ IONICALLY\ SCREENED\ MEDIUM

   Bardeen calculated the sum over bubble diagrams (Fig. 1b).
The correction due to the "bubble"  vertex is very large, and
negative; $i.$ $e.$, for $\omega=0$, the sum over the bubble
diagrams reduces the vertex  from the 
Bloch value 4$\pi Ze^2/q^2$ by a factor of
$1/[1+D_{ee}N(E_F)]$ $\approx$ $1/[1+(q_D/q)^2]$, as in
Eq. (1). For $q\approx k_F$, this factor is typically about 1/3, 
and for small q values, it is even smaller. This reduction 
accounts for the resistivity of monovalent metals being an
order smaller than the value calculated assuming the Bloch
value of the vertex. Now, to include the ionic screening, we
should replace $D_{ee}$ in Fig. 1b by $\tilde{D}_{ee}$.  Here
we must set $q_D =0$ in Eq. (2) to avoid double counting the
electronic screening. Since $\tilde D_{ee}$ is so much smaller
(at low frequencies) than $D_{ee}$, the vertex correction is 
much smaller, and when $\tilde D_{ee}$ can be neglected, 
the vertex becomes the Bloch vertex; i.e. it is considerably 
$\it larger$ . Thus, a very large ionic dielectric constant 
restores the value of the vertex to the large original Bloch
value at low frequencies 
($\omega < \omega_T \approx 20 meV$). To illustrate this
effect we consider $\Gamma _{Bardeen}$ replaced with 
$\tilde\Gamma _{Bardeen}$ where
\begin{equation}
\tilde\Gamma _{Bardeen} = \frac{\Gamma_{Bloch}}
{[1+\frac{D_{ee}N(E_F)}{\varepsilon(0)}]} \approx
\frac{\Gamma_{Bloch}}{[1+\frac{q_D^2}{q^2\varepsilon(0)}]}.
\end{equation}
In Fig. 3 we show $\tilde\Gamma _{Bardeen} / \Gamma _{Bloch}$ 
and $\Gamma _{Bardeen} / \Gamma _{Bloch}$
vs $q/k_F$ for $\varepsilon(0)=40$ and $q_D=k_F/2$. It is clearly
seen that, with increasing q, $\tilde\Gamma _{Bardeen}$ 
approaches $\Gamma_{Bloch}$ much faster then 
$\Gamma_{Bardeen}$ does. At the q values of interest here, 
the corrected vertex is quite close to the original Bloch vertex, 
e.g., for $q=0.2k_F$ we have 
$\tilde\Gamma _{Bardeen} \approx 0.9\Gamma_{Bloch}$.% 
\vspace{0.5in} 

6) CONCLUSIONS

   We have shown that the lowest order vertex corrections including
ionic dielectric effects are large. In particular, we show that the 
electron-phonon vertex for small q and small $\omega$ is greatly 
enhanced over the "standard" Bardeen value. Indeed, surprisingly, 
the Coulomb vertex correction, 
$\Delta\Gamma_{Coulomb}/\Gamma_{Bardeen}$, is
already large. This may be related to the 2D nature of the system.
We are, however, not aware of any explicit calculations of this 
quantity either in 2D or in 3D. It has apparently previously been 
considered phenomenologically to be included in the bare vertex.  
We have also shown that the inclusion of ionic effects also greatly 
affects the electronic screening. These corrections all tend to 
significantly increase the maximum superconducting $T_c$ value. 
This may require us to reconsider the possibility of the high $T_c$ 
as being due, at least partially, to a phonon-mediated interaction. 
At the current level of approximation, however, one cannot make a
definite statement in this respect.

  We point out that we have employed a "hybrid" formalism in
which a 3D $\varepsilon$ is combined with 2d band structure and 
2D integrations. This is not done to make the calculations easier
but to have a consistent picture in accordance with experiment.
We do not believe that a pure 2D or a pure 3D senario is
physically correct in the high-T$_c$ superconductors.
 
   There are still a number of things to be done before
a truly quantitative theory is obtained. We mentioned
briefly in Section 4 that the large contribution of the
lowest order vertex correction indicates that further
vertex corrections may be important, in particular the
particle-hole ladder diagrams that contribute to the
spin susceptibility. At the least, one must consistently
include such diagrams in the bubbles of the electronic
screening diagrams as well as directly as vertex
corrections. If the higher order vertex corrections 
can be summed to infinite order the result could turn
out to be smaller than the large lowest order contribution. 
In any case, this difficult task must be carried out before
quantitative comparison with experiment can be attempted.
In addition, a better theory must go beyond the present 
level of approximation in other respects, for example, 
the momentum dependence of $\Gamma _{Bardeen}$ 
should also be included. More fundamentally, there is 
at present no microscopic, field theoretic
derivation of the Lyddane Sachs Teller expression for 
$\varepsilon$ in Eq.(5), and in general it does not hold for a
polyatomic unit cell. However, experimentally it applies extremely
well for LSCO \cite{Henn} and reasonably well for YBCO 
\cite{Kircher}.
 
We acknolowledge helpful discussions with J. Appel, A. Baratoff,
J. Birman, O. Gunnarsson, V. Kresin, D. Manske, M. Peter, and 
K. Scharnberg. The work of M. W. is supported by the US-Israel BSF.
\vspace{0.5in}  

APPENDIX:

   We display in detail the integrand of Eq.(6).

\begin{eqnarray*}
iI & = &  \frac{(\omega_L^2-\omega_T^2)f(\varepsilon_p)f(\varepsilon_{p+q})}
{(\omega_L-\varepsilon_p)(\omega_L-\varepsilon_{p+q} + q_0 )}
  + 
\frac{(\omega_L^2-\omega_T^2)[1-f(\varepsilon_p)][1-f(\varepsilon_{p+q})]}
{(\omega_L+\varepsilon_p)(\omega_L+\varepsilon_{p+q} - q_0)}
\\
  & + &  
\frac{(\varepsilon_p^2-\omega_T^2)f(\varepsilon_p)[1-f(\varepsilon_{p+q})]}
{(\varepsilon_p - \omega_L)(\varepsilon_{p+q}-\varepsilon_p - q_0 - i\delta)}
  + 
\frac{(\varepsilon_p^2-\omega_T^2)[1-f(\varepsilon_p)]f(\varepsilon_{p+q})}
{(\varepsilon_p + \omega_L)(\varepsilon_{p+q}-\varepsilon_p - q_0)}
\\
  & - & 
\frac{[(\varepsilon_{p+q}-q_0)^2-\omega_T^2][1-f(\varepsilon_p)]f(
\varepsilon_{p+q})}{(\varepsilon_{p+q} -q_0- \omega_L)(
\varepsilon_{p+q}-\varepsilon_p - q_0)}
  - 
\frac{[(\varepsilon_{p+q}-q_0)^2-\omega_T^2]f(\varepsilon_p)[1-f(
\varepsilon_{p+q})]}{(\varepsilon_{p+q} -q_0+ \omega_L)(
\varepsilon_{p+q}-\varepsilon_p - q_0- i\delta)}
\end{eqnarray*}
\vspace{0.5in}  
FIGURE\ CAPTIONS 

1.  The various electron-Boson vertices discussed in the text:
a) The bare (Bloch) vertex; b) The Bardeen vertex which
includes electronic screening. D$_{ee}$ (q) = 4$\pi e^2/q^2$; 
c) The lowest order Coulomb vertex correction; d) The electron-
electron interaction including ionic dielectric effects (Eq.(2)).

2.  The vertex corrections 
$\Delta\Gamma_{Coulomb}/\Gamma_{Bardeen}$ and
$\Delta\Gamma/\Gamma_{Bardeen}$ as functions of the
external momentum q for external frequency
$q_0=0.08E_F$ (solid curves) and $q_0=0.02E_F$ 
(dashed curves).

3.  The electronically screened vertex with 
($\tilde\Gamma _{Bardeen}$) and without 
($\Gamma _{Bardeen}$) ionic screening.


\begin{references}
%
%
\bibitem{Zhang} S. C. Zhang, Science {\bf 285}, 1089 (1997].
%
\bibitem{Birman} J. A. Birman and A. I. Solomon, 
Phys. Rev. Lett. {\bf 49}, 230 (1982).
%
\bibitem{Weger_etal} M. Weger, M. Peter, and L. P. Pitaevskii, 
Z. Phys. B {\bf 101}, 573 (1996).
%
\bibitem{Peter} M. Peter, M. Weger, and L. P. Pitaevskii,
Ann. Phys. {\bf 7}, 174 (1998).
%
\bibitem{Bardeen} J. Bardeen, Phys. Rev. {\bf 52}, 688 (1937).
%
\bibitem{Kircher} J. Kircher, R. Henn, M. Cardona, P. L. Richards, 
and G. P. Williams, J. Opt. Soc. Am. B {\bf 14}, 705 (1997).
%
\bibitem{Henn} R. Henn, A. Wittlin, M. Cardona, and S. Uchida,
Phys. Rev. B {\bf 56}, 6295 (1997).
%
\bibitem{Pietr} L. Pietronero, S. Straessler, and C. Grimaldi,
Phys. Rev {\bf B 52}, 10516 (1995).
%
\bibitem{Bloch} F. Bloch,  Z. Phys. {\bf 52}, 555 (1928).
%
\bibitem{Allen} P. B. Allen, M. L. Cohen, and D. R. Penn,
Phys. Rev. {\bf B 38}, 2513 (1988).
%
\bibitem{Gersten} J. I. Gersten and M. Weger, Physica {\bf B 225}, 
33 (1996).
%
\bibitem{Sham} M. Grabowsky and L. J. Sham, Phys. Rev. {\bf B 29}, 
6132 (1984).
%
\bibitem{Santi} G. Santi, T. Jarlborg, M. Peter, and M. Weger,
Physica {\bf C 259}, 253 (1996).
%
\bibitem{Varelog} A. Perali and G. Varelogiannis, Phys. Rev. {\bf B 61},
3672 (2000).
%
\bibitem{Abrik} A. A. Abrikosov, Physica {\bf C 244}, 243 (1995).
%
\bibitem{Zeyher} R. Zeyher and M. L. Kulic, Phys. Rev. {\bf B 53}, 
2850 (1996).
%
\bibitem{Bulut} N. Bulut and D. J. Scalapino, Phys. Rev. {\bf B 54}, 
14971(1996).
%
\bibitem{Tsuei} C. C. Tsuei and J. R. Kirtley, Chin. J. Phys. (Taiwan) {\bf 36},
149 (1998).
%
\bibitem{Leggett} A. J. Leggett, Phys. Rev. Lett. {\bf 83}, 392 (1999).
%
\bibitem{Bianconi} A. Bianconi $et$ $al,$ Phys. Rev. Lett. {\bf 76}, 
3412 (1996); Europhys. Lett. {\bf 31}, 411 (1995); J. M. Tranquada,
in CP483, {\it High Temperature Superconductivity}, (eds. S. E. Barnes,
J. Ashkenazi, J. L. Cohn, and F. Zuo, 1999 AIP), p. 336.
%
\bibitem{Rome} M. Weger, J. Superconductivity {\bf 10}, 435 (1997).
%
\bibitem{Appel} J. Appel, Phys. Rev. {\bf 180}, 508 (1969). A soft mode, 
transforming according to the identity representation of the tetragonal
point group, is of particular interest for the superconductivity of
SrTiO$_{3-x}$. See also G. Binnig et al, Phys. Rev. Lett. {\bf 45},
1352 (1980).
%
\bibitem{LST} N. W. Ashcroft and N. B. Mermin, $Solid State Physics$, 
(Holt Saunders Philadelphia 1976), pp. 547-548.
%
\bibitem{Vedeneev} S. I. Vedeneev, P. Samuely, S. V. Meshkov,
G. M. Eliashberg, A. G. M. Jansen, and P. Wyder,   Physica C {\bf 198}, 
47 (1992).
%
\bibitem{Kamimura} H. Kamimura and A. Sano, J. Superconductivity 
{\bf 10}, 279 (1997).
%
\bibitem{Anisimov} V.I. Anisimov, M. A. Korotin, J. Zaanen, 
and O. K. Anderson, Phys. Rev. Lett. {\bf 68}, 345 (1992);
V. Meregalli and S. Y. Savrasov, Phys. Rev. B {\bf 57},
14453 (1998).
%
\bibitem{Polinger} V. Polinger, D. Haskel, E. A. Stern, cond-mat/9811425.
%
\bibitem{Gatt} R. Gatt $et$ $al,$ cond-mat/9906070
% 
\bibitem{Manske} D. Manske, C. T. Rieck, and D. Fay, J. Low Temp.
Physics {\bf 99}, 527 (1995); Physica {\bf C 235-240}, 2129 (1994).
%
\bibitem{Kim}D. J. Kim, Physics Reports {\bf 171}, (1987)129 (1987).
%
\bibitem{McMill}W. L. McMillan, Phys. Rev. B {\bf 167}, 331 (1968).
%
\bibitem{Heine} V. Heine, P. Nozieres, and J. W. Wilkins, Phil. Mag.
{\bf 13}, 741 (1966).
%
\bibitem{Roehler} J. R\"ohler, J. Superconductivity {\bf 9}, 457 (1996).
%
%
\end{references}
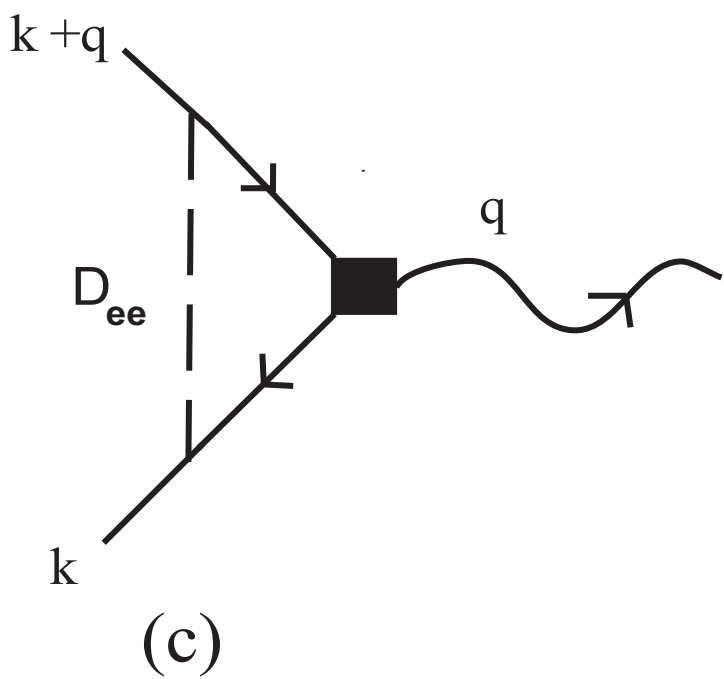
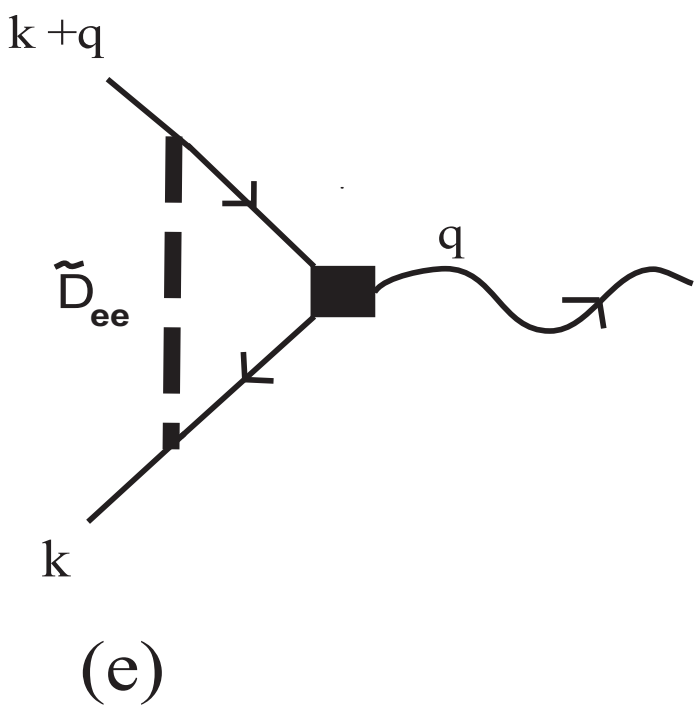
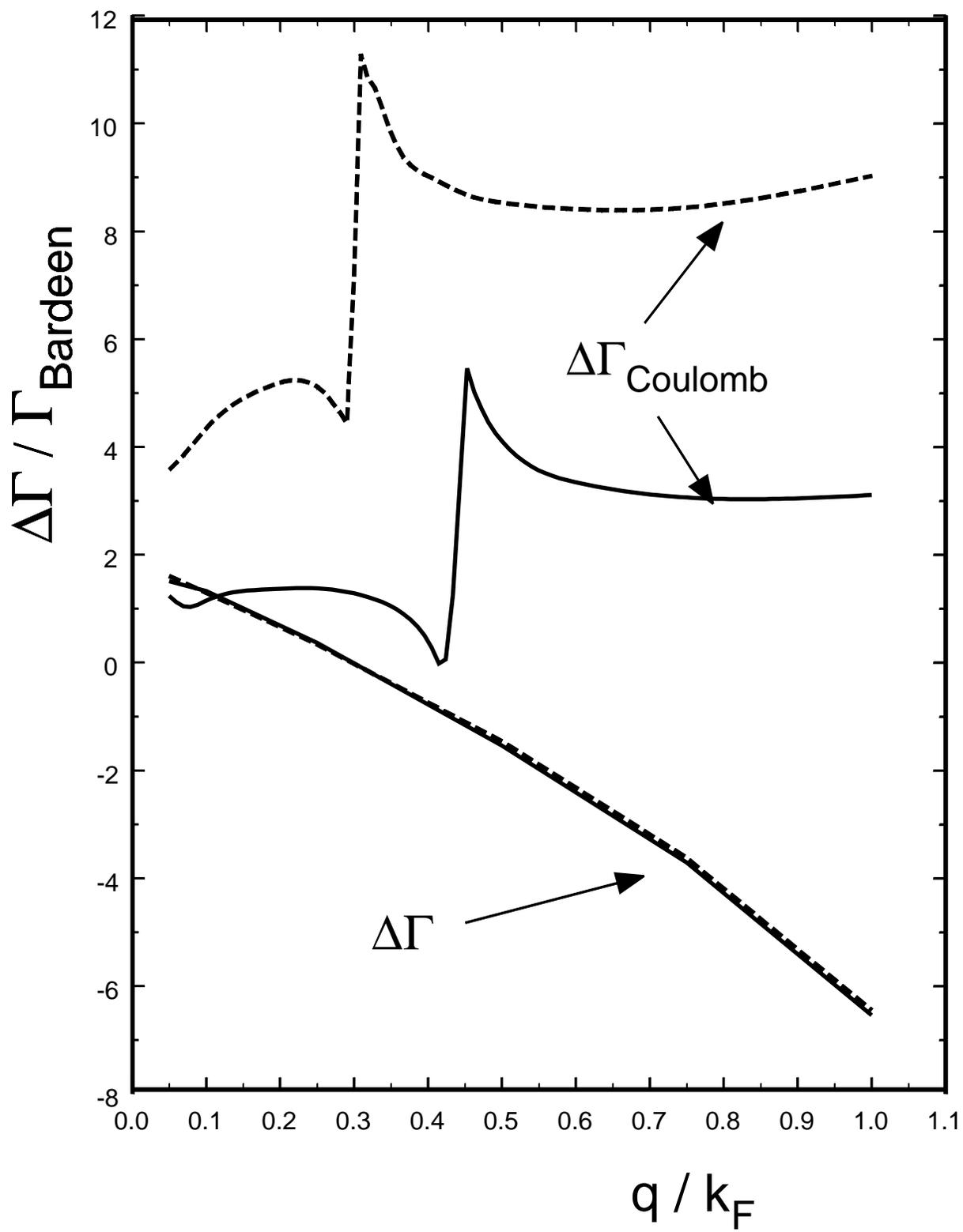
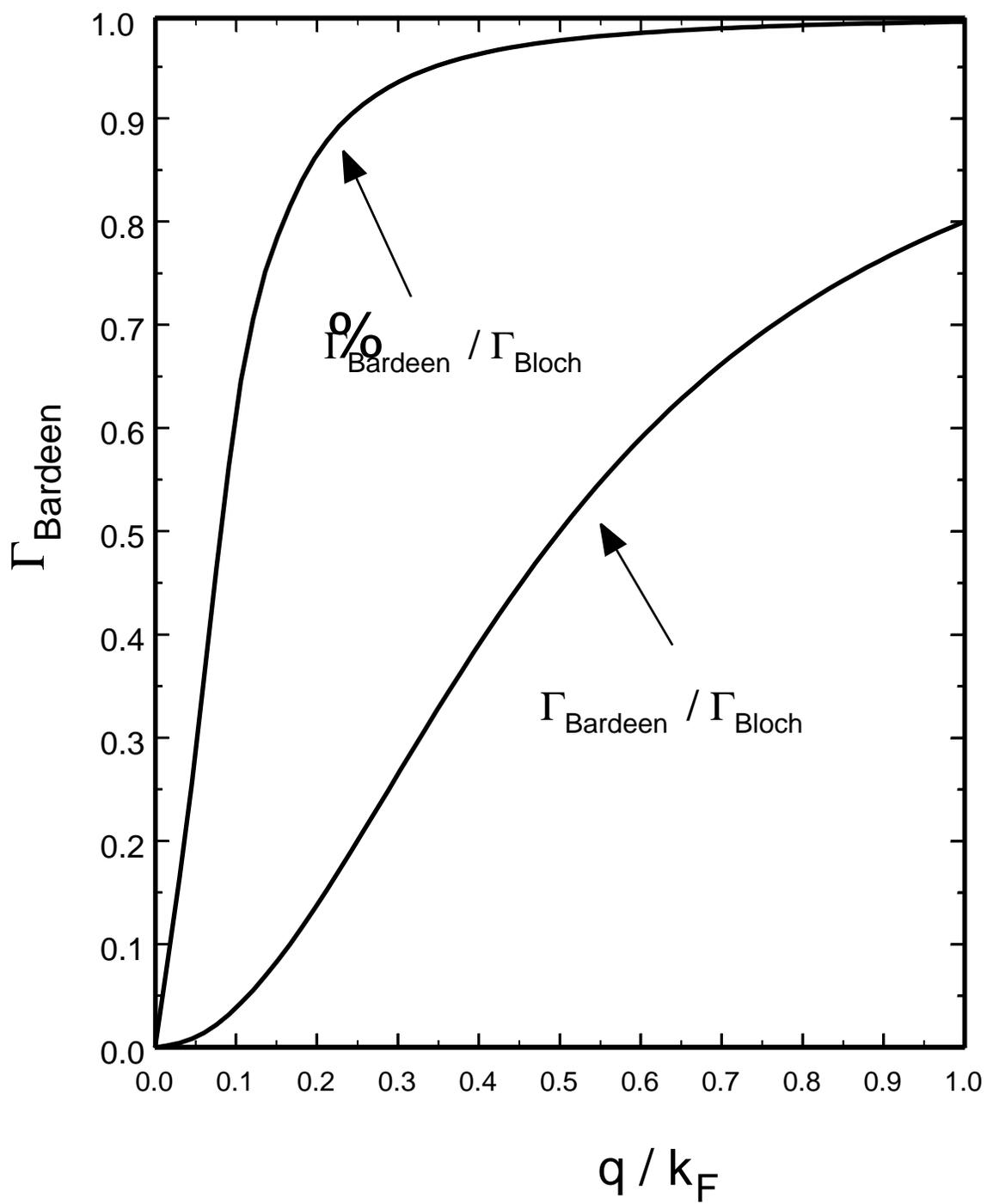
\end{document}